\documentclass[twocolumn,showpacs,prl]{revtex4}
\usepackage{dcolumn}
\usepackage{bm}

\begin{document}

\title{Probing the Higgs Field Using Massive Particles as Sources
and Detectors}

\author{S. Reucroft$^1$, Y.N. Srivastava$^{1,2}$, \\
J. Swain$^1$ and A. Widom$^1$}

\affiliation{1. Physics Department, Northeastern University \\
110 Forsyth Street, \\
Boston, MA 02115, USA\\
2. Physics Department and INFN\\
University of Perugia,\\
Perugia, Italy\\
E-mail: john.swain@cern.ch}

\begin{abstract}
In the Standard Model, all massive elementary particles acquire
their masses by coupling to a background Higgs field with a
non-zero vacuum expectation value. What is often overlooked  is
that each massive particle is also a source of the Higgs field. A
given particle  can in principle shift the mass of a neighboring
particle.  The mass shift effect goes beyond the usual
perturbative Feynman diagram calculations which implicitly assume
that the mass of each particle is rigidly fixed. Local mass shifts
offer a unique handle on Higgs physics since they do not require
the production of on-shell Higgs bosons. We provide theoretical
estimates showing that the mass shift effect can be large and
measurable, especially near pair threshold, at both the Tevatron
and the LHC.
\end{abstract}

\maketitle

\section{Introduction \label{intro}}

In the usual treatments of a Poincar\'{e} invariant field theory,
particles are labelled according to irreducible representations of
the space-time symmetry group, and labelled according to values of
two Casimir invariants which can be constructed from spin and mass
\cite{Weinberg}. Within this framework, it is common practice in
high energy physics to view the mass of a particle, just as its
spin, as an intrinsic and immutable property completely unaffected
by its surroundings.


That being said, in the Standard Model \cite{PeskinandSchroder}
``mass'' is more than just a representation label with no  more
dependence on any external fields than the spin of a particle. In
fact, all of the usual {\em elementary} fermions and bosons
(except for the Higgs boson itself) are, in the absence of
interactions, {\em massless}. That is, there is no single
non-dynamical parameter that appears as a mass  term for any of
the fermions or gauge bosons in the Standard Model Lagrangian.
Rather, there is  an emergent mass through a coupling to a scalar
field whose dynamics have been arranged for it to have a non-zero
value which is independent of space and time ({\em i.e.} to
preserve the Poincar{\'e} symmetry of space-time). Couplings to
this  field, assumed constant in space and time due to the
dynamics of the theory, play the roles of masses.

Of course actual experiments need not be Poincar\'{e} invariant,
and in general the presence of an experimental setup will break
this invariance. In condensed matter physics, one often needs to
consider the effects of an environment, and a new mass (shifted by
electromagnetic interactions with matter, for example, and not
necessarily even a scalar any longer) must often be introduced
\cite{Condmat},  together with a different symmetry group. Even
outside bulk condensed matter, electromagnetic mass shifts are
commonly introduced in the literature due either to external
fields \cite{LandauandLifschitz,vanHolten} or even to changes in
the vacuum fluctuations due to boundaries \cite{Barton}. In
nuclear physics similar phenomena occur. For example, a free
neutron is unstable but, when bound to a proton in a deuteron, it
becomes stable since it is then effectively too light to decay. In
each case one can think of mass as being due to two parts: one
somehow ``intrinsic'' and one due to interaction with an external
field.

As stated earlier, in the Standard Model \cite{PeskinandSchroder}
these ideas are taken to an extreme so that even in the vacuum
{\em all} of the mass of the fermions and gauge bosons is due to
interaction with an external field - the Higgs field - which is
taken to have a non-vanishing vacuum expectation value.

While for many applications the background Higgs field can be
considered a space-time constant, the Standard Model asserts that
the Higgs field is a dynamical object. Indeed, if it were not,
then there would be no way to tell that it exists at all!
Approaches so far \cite{EWWGH} to detecting the Higgs field have
concentrated on  looking for its quanta: Higgs bosons. One hope is
that accelerators  will have enough energy to produce a particle
on shell, but so far all  that has been done is to rule out a
Higgs with a mass below 114.4 GeV/$c^2$ at 95 \%
confidence level.

Another approach is to look for measurable quantities involving
Higgs boson exchange between heavy particles that could reveal its
existence via radiative corrections. A recent best fit result
\cite{Higgsradiative}  from radiative corrections gives, with
large errors, a Higgs mass of 168 GeV/$c^2$  for an assumed top
quark mass of 178 GeV/$c^2$, placing detection of an on-shell
Higgs boson likely out of reach of the Tevatron at Fermilab and
leaving its  discovery to the LHC at CERN. The LEP Electroweak
Working Group makes its most recent analyses of combined data
available at \cite{EWWG}.  Their preferred value for the Higgs
mass is 129 GeV/$c^2$, with an experimental uncertainty of +74 and
-49 GeV/$c^2$ (at 68\% confidence level, not including theoretical
uncertainty). This rather large and asymmetric uncertainty is due
to the fact that in typical radiative corrections the Higgs mass
appears only logarithmically so in fact even quite high Higgs
masses of hundreds of GeV/$c^2$ are not ruled out.

Fortunately, the notion of a dynamical Higgs field leads one to another
approach that has not yet been suggested to the best of our knowledge:
to probe the Higgs field not directly in terms of its quanta, but rather as
a field which changes masses. The field is sourced to a significant extent
by heavy particles (ones which couple strongly to the Higgs
field) and the resulting source-modified Higgs field  might then be detected
by {\em other} heavy particles via induced mass shifts.

The {\em static} Higgs field $\sigma$ at a spatial point ${\bf r}$,
\begin{math} \sigma ({\bf r}) \end{math},
produced by a particle of mass \begin{math} m \end{math}
fixed at rest on the coordinate origin is given by
\begin{equation}
\sigma ({\bf r})=-\left(\frac{mc}{4\pi \hbar v}\right)
\frac{ e^{-m_Hcr/\hbar }}{r} ,
\label{intro1}
\end{equation}
where the vacuum expectation value of the Higgs field without the source
is
\begin{math} v=\left<\phi \right>  \end{math}, the total Higgs field is
\begin{math} \phi=v+\sigma  \end{math}  and \begin{math}  m_H \end{math}
is the Higgs mass. The rapid falloff with distance for a Higgs boson
mass of a few hundred GeV/$c^2$, together with the
weakness of the Higgs coupling to all but the most massive particles makes
it difficult to suggest suitable laboratory experiments to examine
Higgs-induced  mass shifts. One hopeful situation is the production of one
massive particle together with a second massive particle. Let us consider
this case in detail.

\section{The Mass of a Particle in the Presence of Another \label{massshift3}}

The source of the Higgs field is the trace of the energy-pressure tensor
which may be formally computed by differentiating the lagrangian density
with respect to the elementary particle masses of the model
\begin{equation}
T^\mu _{\ \ \mu }(x) \equiv  T(x)=\sum_a m_a \frac{\partial \mathcal{L}(x)}
{\partial m_a} .
\label{source1}
\end{equation}
Thus, a fermionic or bosonic source of the Higgs field would have the form
\begin{eqnarray}
T_{\rm fermion}(x) &=& -c^2 m_F\bar{\psi}(x)\psi(x),
\nonumber \\
T_{\rm boson}(x) &=& -\left(\frac{m_B^2c^3}{\hbar}\right)\bar{B}(x){B}(x).
\label{source2}
\end{eqnarray}
For a {\em classical particle} with the proper time action
\begin{math}  S_{\rm classical}=-mc^2\int d\tau   \end{math}, the lagrangian density
\begin{math}
\mathcal{L}_{\rm classical}(x) =-mc^3\int \delta\big(x-x(\tau )\big)d\tau
\end{math}
yields the classical source
\begin{equation}
T_{classical}(x)=-mc^3\int \delta\big(x-x(\tau )\big)d\tau .
\label{source3}
\end{equation}

The idea now is very simple. Consider a massive particle ``$1$''
of mass $m_1$ adjacent (in a space-time picture) to another
particle ``$2$'' of mass $m_2$. Here $m_1$ and $m_2$ refer to
their masses in the usual sense of a Yukawa coupling times the
background Higgs vacuum expectation value. The claim is that
particle 1 will couple to the Higgs field produced by fermion 2
and have its mass shifted by an amount proportional to its own
mass $m_1$ (its coupling to the Higgs fields) and also
proportional to $m_2$ (the strength of particle 2's coupling to
the Higgs field). The relevant Higgs mass shift coupling strength
may be written
\begin{equation}
\alpha_H= \frac{c^2 m_1m_2}{4\pi \hbar^2v^2}
=\frac{\sqrt{2}G_Fm_1m_2}{4\pi \hbar c}
\label{source4}
\end{equation}
where \begin{math} G_F  \end{math} is the Fermi coupling strength.  The
relevant energy scale is
\begin{math} (\hbar v/c)\approx 246\ {\rm GeV}  \end{math}
so that only heavy particle pairs, e.g.
\begin{math} W^+W^-,\  ZZ, \end{math} or \begin{math} \bar{t}t  \end{math},
have an appreciable mutual coupling strength.

It turns out that the mass shift is only weakly dependent on the
Higgs particle mass in that the light cone singularity of the
Higgs propagator for neighboring events is mass independent.  No
real (on-shell) Higgs boson needs to be produced for the mass
shift any more than a real photon needs to be produced to provoke
an electromagnetic Lamb energy shift, or a real pion needs to be
produced to make a neutron in a deuteron stable.  In detail, the
propagator
\begin{equation}
D(x-y)=\int \left[\frac{e^{ik\cdot (x-y)}}
{k^2+\kappa^2 -i0^+} \right]
\frac{d^4k}{(2\pi )^4}\ ,
\label{soursce5}
\end{equation}
(where \begin{math} \ \hbar \kappa=m_Hc  \end{math})
determines the Higgs field at particle 2 due to particle 1 as given by
\begin{equation}
\sigma_2(x)=\frac{1}{\hbar cv}\int D(x-y)T_1(y)d^4y.
\label{source6}
\end{equation}
For example if particle 1 moves on a path
\begin{math} x_1(\tau_1)  \end{math}, then Eqs.(\ref{source3})
and (\ref{source6}) imply
\begin{equation}
\sigma_2(x)=-\frac{m_1c^2}{\hbar v}\int D\big(x-x_1(\tau_1)\big)d\tau_1.
\label{source7}
\end{equation}
If particle 2 moves on a path
\begin{math} x_2(\tau_2)  \end{math}, then
the added action to particle 2 due to particle 1 is
\begin{eqnarray}
S_{21} &=& \frac{1}{cv}\int T_2(x)\sigma_2(x)d^4x,
\nonumber \\
S_{21} &=& -\frac{m_2c^2}{v}\int \sigma_2\big(x_2(\tau_2) \big)d\tau_2 ,
\nonumber \\
S_{21} &=& \frac{m_1m_2 c^4}{\hbar v^2}
\int \int D\big(x_1(\tau_1)-x_2(\tau_2)\big)d\tau_1 d\tau_2\ .
\label{source8}
\end{eqnarray}
Here it is of use to recall the Feynman -Wheeler formulation of electrodynamics
in which the interaction between two point charges has the form
\begin{equation}
S_{21}({\rm photon})=\frac{e_1e_2}{c}\int_{P_1}\int_{P_2}D_{\mu \nu }
(x_1-x_2)[dx^\mu _1dx^\nu _2].
\label{source9}
\end{equation}
$P_1$ is the path of charge 1,  $P_2$ is the path of charge 2
and $D_{\mu \nu}$ is the photon propagator.
The Higgs exchange analog to the Feynman-Wheeler interaction has
been derived in Eq.(\ref{source8}); It is
\begin{equation}
S_{21}=\frac{\sqrt{2} G_F m_1m_2}{c}\int_{P_1}\int_{P_2}
D(x_1-x_2)[cd\tau_1 cd\tau _2].
\label{source10}
\end{equation}
To compute the mass shifts for the two particles due to their
mutual interactions
when mass $m_1$ travels on path $P_1$ and mass $m_2$ travels along path
$P_2$ one need only apply the rule
\begin{equation}
{\Re e}\big(S_{21}\big)=- c^2\Delta m_1\int_{P_1} d\tau_1
=- c^2\Delta m_2\int_{P_2} d\tau_2.
\label{source11}
\end{equation}
In particular, the mass shift in particle 2 due to the Higgs field
produced by particle 1 is given by
\begin{eqnarray}
\Delta m_2 &=&-\left(\frac{\sqrt{2} G_F m_1m_2}{c}\right)\times
\nonumber \\
&\ & \frac{\int_{P_1} \int_{P_2} {\Re e}D\big(x_1-x_2\big)d\tau_1 d\tau_2}
{\int_{P_2} d\tau_2}\ .
\label{source12}
\end{eqnarray}
Suppose that particles 1 and 2 have the four momenta $p_1=m_1v_1$ and
$p_2=m_2v_2$ (where $v_1$ and $v_2$ are four-velocities)
and thus the invariant mass $\sqrt{{s}}$ as given by
\begin{eqnarray}
-c^2{s} &=& (p_1+p_2)^2 ,
\nonumber \\
-\left(\frac{v_1\cdot v_2}{c^2}\right) &=&
\frac{{s}-(m_1^2+m_2^2)}{2m_1m_2}\ .
\label{source13}
\end{eqnarray}
The real part of the Higgs propagator ${\Re e}D(x-y)$ vanishes if $x$ and $y$
are space-like separated. If $x$ and $y$ are not space-like separated, then
${\Re e}D(x-y)$ has two terms: (i) There is a light-cone singularity which
is independent
of the Higgs mass. (ii) There is a finite smooth portion which depends on
the Higgs mass
$m_H=(\hbar \kappa /c)$. In terms of the first order Bessel function
$J_1(\xi )$ we have
\begin{eqnarray}
{\Re e}D(x) &=& 0\ \ \ \ {\rm for\ space-like}\ \  \  x^2>0,
\nonumber \\
{\Re e}D(x) &=&\frac{1}{4\pi }\left[\delta (x^2)-
\frac{\kappa J_1\big(\kappa \sqrt{-x^2}\big)}{2\sqrt{-x^2}}\right]
\ \  x^2\le 0.\ \ \ \ .
\label{source14}
\end{eqnarray}
The light-cone singularity dominates the mass shift in Eq.(\ref{source12}).

The proper time integral lasts (on average) as long as the particle life-time
$\int d\tau_2 =\Gamma_2^{-1}$  so that the light-cone singularity
approximation
in Eq.(\ref{source12}) reads
\begin{eqnarray}
\frac{\Delta m_1}{\Gamma_1}
&\approx &-\left(\frac{\sqrt{2} G_F m_1m_2}{4\pi c}\right) \times
\nonumber \\
&\ &\int \int \delta \big((v_1\tau_1-v_2\tau_2)^2\big)d\tau_1 d\tau_2,
\label{source15}
\end{eqnarray}
where ${4\pi \Re e}D(x_1-x_2)\approx \delta \big((x_1-x_2)^2\big)$ has been
invoked and, of course,
$(\Delta  m_1/\Gamma_1)=(\Delta  m_2/\Gamma_2)$.
The double integral on the right hand side of Eq.(\ref{source15})
has a logarithmic singularity of the form
\begin{eqnarray}
& \  &c^2\int \int
\delta \big((v_1\tau_1-v_2\tau_2)^2\big)d\tau_1 d\tau_2 \approx
\nonumber \\
&\ & \ \ \ \ \
\frac{1}{\sqrt{(v_1\cdot v_2/c^2)^2-1}}
\ln \left(\frac{\tau_{max}}{\tau_{min}}\right).
\label{source16}
\end{eqnarray}
The maximum and minimum proper times ($\tau_{max}$ and
$\tau_{min}$) must now be estimated, but as they only appear
logarithmically, our results depend only weakly on how this is
done. The maximum proper time $\tau_{max}$ is determined by
particle life-times $\tau_{max}\sim \Gamma^{-1}$. The minimum
proper time is determined by the duration of the classical path
viewpoint $\tau_{min}\sim (\hbar/mc^2)$. We then estimate $(\Delta
m_1/\Gamma_1)$ as
\begin{eqnarray}
\left(\frac{\Delta m_1}{\Gamma_1}\right)&\approx &
 -\left(\frac{\sqrt{2}G_F m_1m_2}{4\pi c^3}\right)\times
\nonumber \\
&\ & \frac{1}{\sqrt{(v_1\cdot v_2/c^2)^2-1}}
\ln \left[\frac{c^2(m_1+m_2)}{\hbar(\Gamma_1+\Gamma_2)} \right] .
\ \ \ \ \ \
\label{source17}
\end{eqnarray}
In terms of the invariant mass $\sqrt{{s}}$, Eq.(\ref{source17}) reads
\begin{eqnarray}
 \frac{c^2\Delta m_1}{\hbar \Gamma_1}
&\approx& -\left(\frac{c^2 m_1m_2}{2\pi \hbar^2v^2}\right)
\nonumber \\
&\ &
\times\sqrt{\frac{m_1^2m_2^2}{{s}^2-2{s}(m_1^2+m_2^2)+(m_1^2-m_2^2)^2}}\
\nonumber \\
&\ &
\times\ln \left[\frac{c^2(m_1+m_2)}{\hbar(\Gamma_1+\Gamma_2)} \right] .
\label{source18}
\end{eqnarray}
Eq.(\ref{source18}) is the central theoretical result of this work.
As can easily be seen, at least near threshold, the right hand side
is the product of terms of order unity so the effect need not be small!

An immediate consequence of the predicted mass shifts is, of
course, also a change in widths. To a good leading approximation,
most heavy particles of interest such as $t$, $W$, $Z$ decay
mainly into two bodies. In these cases the phase space is
proportional to the 3-momenta of the outgoing particles. As is
well-known, this is proportional to
$\sqrt{\lambda(m_A^2,m_B^2,m_C^2)}/m_A$ where $A$ represents the
heavy particle, $B$ and $C$ its decay products, and
$\lambda(x,y,z)= x^2+y^2+z^2-2xy-2yz-2zx$. In most cases (the main
exception being $t\rightarrow Wb$) the outgoing particles can be
considered almost massless and the width of a $Z$ or $W$ is
proportional to its mass. In either case, the change $\Delta
\Gamma$ in width $\Gamma$ is easily obtained and seen to be large,
{\em i.e.} $(\Delta \Gamma/ \Gamma) \approx (\Delta M/ M) $.

\section{Experimental Comments\label{experiment}}

The key experimental message that we wish to communicate in this
paper is that if one is to produce heavy particles (where
``heavy'' means with mass not small compared to 246 GeV/$c^2$) in
conjunction with other heavy particles, there is good information
in the measured  distribution of masses and widths as a function
of relative velocities (or center of mass energy) as these need
not be the same as one would expect from single production. This
gives information on the structure of the Higgs field produced by
heavy particles {\em even if there is not enough energy to produce
an on-shell Higgs boson}. In particular, this means that
kinematical fits using as input masses obtained from other
experiments, where heavy particles are singly produced, should not
be done without great care. It also means that mass cuts and other
kinematical cuts obtained from experiments where heavy particles
are singly produced, are not necessarily reliable if other heavy
particles are produced in association.

A concrete example of how data might be approached is perhaps in
order. Suppose one is looking at the production of $Z$-boson
pairs. We take this example since the
many other concerns about final state
interactions can be neglected: there are no one-gluon or
one-photon exchange potentials and Yukawa potential effects
\cite{earlierpaper}. If each $Z$ boson
decays into $e^+e^-$ or $\mu^+\mu^-$ one has access to the full
kinematics in a rather clean environment. With the invariant
masses of  $e^+e^-$ and $\mu^+\mu^-$ plotted as a function of
relative velocity or center of mass energy
between the $Z$'s, and with enough statistics,
the predicted mass shift could be detectable.

Even in the absence of a more reliable theoretical estimate of all
factors involved at this moment, the data are certainly worth
looking at with an open mind (and relaxed cuts and no kinematical
mass fits).

There are several processes that can be investigated
experimentally and that have the potential to see the effect
described in this paper. These are $Z^0Z^0$ versus single $Z^0$
production, $W^+W^-$ versus single $W^{\pm}$ production and
$t\overline{t}$ versus single $t$ production. In addition, a
measurement of the mass of these particles near pair threshold can
be compared to the mass when the particles are far from pair
threshold.

The mass of the $Z^0$-boson has been determined very precisely at
LEP1 \cite{ALEPHZ,DELPHIZ,L3Z,OPALZ} yielding the result
\cite{PDG} $M_Z = 91.1876 \pm 0.0021\ {\mathrm{GeV}}/c^2$. These
are all measurements made at the $Z^0$ pole. All four LEP
experiments saw clear $Z^0$ signals at LEP2, but none made a
separate $Z^0$ mass determination
\cite{ALEPHZZ,DELPHIZZ,L3ZZ,OPALZZ}. All four experiments are
consistent with $M_Z$ being independent of production mechanism
(with an uncertainty of $\lesssim$ 1\%); it is clear from the
cross-section data that the $Z^0Z^0$ threshold is close to 180
GeV.

The mass of the $W^\pm$-boson has been determined quite precisely
at LEP2 \cite{ALEPHWW,DELPHIWW,L3WW,OPALWW} and at the Tevatron
collider \cite{CDFWW,DZEROWW}. The average of all these
measurements \cite{EWWG, PDG} gives $M_W = 80.425 \pm 0.034 \
{\mathrm{GeV}}/c^2$.

In addition, indirect determinations of the $W$ mass have been
made. One of these is from a careful measurement of
$\sin^2\theta_W$  by the NuTeV collaboration \cite{NUTEVW} and,
assuming the value of $M_Z$ from LEP1, gives $M_W = 80.136 \pm
0.084 \ {\mathrm{GeV}}/c^2$. The LEP Electroweak Working Group has
also determined $M_W$ from a global standard model fit to the SLD
data, LEP1 data and the best measurement of $M_t$ \cite{EWWG}.
They quote $M_W = 80.373 \pm 0.023 \ {\mathrm{GeV}}/c^2$.

The $t$ mass has been determined by CDF \cite{cdft} and D\O\
\cite{dzerot} and the combined average value at the time of
writing \cite{combinedtop} is $m_t = 172.7 \pm 2.9 \
{\mathrm{GeV}}/c^2$. The $t$'s are presumably produced in pairs.
There is no determination to date of $m_t$ in an environment where
the $t$ is produced alone, although such a measurement is
important because the $t$-quark could potentially provide the most
sensitive probe of the Higgs field. The current status of top
quark measurements from the Tevatron Electroweak Working Group can
be found at \cite{tevewwg}. It would be interesting to investigate
very carefully the events where the $t\overline{t}$ effective mass
is close to threshold.

While the data published to date do not contain enough information
on the event kinematics in order to establish the effect we
describe here, we hope that this paper will stimulate further
analyses. In particular, it makes sense to study the masses
determined for particles produced in pairs near and far from
threshold, or even better, as a function of center-of-mass energy.
Single production and pair production well above threshold should
give masses in good agreement with each other, and corresponding
to the usual notion of the ``mass'' of a particle. Masses obtained
from pairs near threshold could be significantly lower.

Of course there is always danger in attempting to reanalyse or
reinterpret published data in light of a new way of thinking about
the analysis, and references here to published data are meant only
to describe the current state of the art and not to claim evidence
for or against the predicted effect. Data are always analysed with
certain theoretical expectations in place and they can affect
published results in ways that are impossible to judge without
access to the original data. For example, one might well reject
candidate $Z^0Z^0$ events on the basis of a low reconstructed $Z$
mass relative to expectations from singly-produced $Z$ bosons at
LEP, while in fact such events could show evidence for a
Higgs-induced mass reduction! These considerations are of even
greater importance for heavier particles such as the top quark and
anything still heavier but not yet discovered that might be
produced at future accelerators.

\section{A Note on Mach's Principle}

Finally, it is interesting to note the distinctly Machian nature
of this result: the mass of a particle is due, at least in part,
to its interactions with all other particles. This reflects a
greater degree of background independence of the Standard Model
\cite{Smolin} than is usually considered, since not only the
background Higgs field but in fact all the local masses are to
some extent dynamically determined. The fact that masses are {\em
reduced} due to interactions with a scalar (and thus attractive)
field produced by surrounding particles would seem, however, to
offer little hope for a Machian picture of a scalar interaction
being responsible for the inertial mass of an object which would
be thought of as intertialess in an empty universe. A similar
conclusion might seem to follow for the attractive spin-2 force
associated with gravitation, however this line of reasoning lies
firmly within linearised general relativity and requires more
careful consideration.

\section{Conclusions\label{conclusions}}

The Standard Model predicts that particles not only obtain their masses
from coupling to a background Higgs field, but that they themselves are
the sources of a Higgs field which can modify the masses of nearby
particles. While a full calculation contains many subtleties, its sign is
unambiguous, and reasonable estimates of the effect in the production of
pairs of heavy particles are that the effect can be very large, especially
near threshold. An additional feature of the effect which is quite
appealing in terms of whether the Higgs mechanism is indeed responsible
for mass or not is that the predicted effect, at least close to threshold
in pair production, is largely independent of Higgs mass. In other
words, failure to observe the effect could rule out a Higgs boson of
any mass at all, even well-beyond the reach of the LHC after many years
of running. On the other hand, observation of the effect would lend
strong direct experimental support to the existence of a Higgs boson
while leaving its actual mass largely undetermined without a careful
study of the center-of-mass energy dependence of the effect.

While completely within the Standard Model, this new effect is beyond the
usual perturbative Feynman diagram calculations and thus, although
straightforward physically, seems to have escaped notice so far. The
$W^\pm$ and $Z^0$ bosons have both been observed in environments where
they are produced singly or in pairs and could offer some information on
the Higgs sector of the Standard Model of a nature different from direct
searches for on-shell particles and radiative corrections assuming fixed
particle masses. Future analyses with top quarks offer even more
information.

Experimental analyses invariably make assumptions about the nature of
what is being observed. Now that the case has been made that masses --
until now thought to be independent of production mechanism  -- may in
fact vary, the possibility of new information from old data begins to
open up. One simple fact which already may be in conflict with published
data  so far, albeit at low statistical significance, is that on general
grounds, one expects that heavy particles produced in pairs will be less
massive than ones which are singly produced. Just how much, of course,
depends on kinematical details which are not easy to extract from
published data.

The experimental situation is both tantalizing in light of data
which exist now, and very promising with the expectation of more
relevant data both from the Tevatron and the LHC. Of course a 
high luminosity linear collider which could scan the energy regions
of interest near threshold for pair production of various heavy
particles would also be of great interest.

\section{Acknowledgements}

We would like to thank our colleagues on LEP, Tevatron and LHC
experiments, and the NSF and INFN for their continued and generous
support.

\end{document}